\documentclass[12pt,aps,showpacs,eqsecnum,nofootinbib,floatfix]{revtex4}
\usepackage{amsmath}
\usepackage{amsfonts}
\usepackage{amssymb}
\usepackage{color}
\usepackage{graphicx}
\usepackage{mathrsfs}

\def\bc{\begin{center}}
\def\nno{\nonumber}
\def\ec{\end{center}}
\def\be{\begin{eqnarray}}
\def\ee{\end{eqnarray}}

\def\dS{dS}
\definecolor{dyellow}{rgb}{1.,0.8,.0}
\definecolor{myblue}{rgb}{.1,.1,.7}
\definecolor{dcyan}{rgb}{.0,.6,.6}
\definecolor{dmagenta}{rgb}{0.6,0.0,0.6}
\definecolor{brown}{rgb}{0.6,0.2,0.}
\definecolor{darkblue}{rgb}{.0,.0,0.5}
\definecolor{darkred}{rgb}{0.75,0.0,0.0}
\definecolor{orange}{rgb}{1.,.6,.0}
\definecolor{dorange}{rgb}{0.8,.4,.0}
\definecolor{darkgreen}{rgb}{0.0,0.6,0.0}
\definecolor{purple}{rgb}{.4,.0,.4}
\definecolor{lightgrey}{rgb}{0.7, 0.7, 0.7}
\definecolor{grey}{rgb}{0.4, 0.4, 0.4}
\def\black{\color{black}}

\def\yellow{\color{yellow}}

\def\dl{\delta}

\def\ka{\kappa}
\def\la{\lambda}
\def\th{\theta}
\def\si{\sigma}
\def\om{\omega}

\def\La{\Lambda}

\def\Om{\Omega}

\def\d#1#2{\frac{\displaystyle #1}{\displaystyle #2}}

\newcommand{\vect}[1]{\mbox{\boldmath $#1$}}

\newcommand\btd{\raise 2pt
\hbox{$\hat\bigtriangledown$}\hskip 1.5pt}
\newcommand\bt{\raise 2pt
\hbox{$\bigtriangledown$}\hskip 1.5pt}

\newcommand{\omits}[1]{}

\def\PRD{{Phys. Rev.}~{\bf D}}
\def\PRL{{Phys. Rev. Lett. }}
\def\PLA{{Phys. Lett.}~{\bf A}}

\begin{document}

\title{De Sitter spacetimes with torsion
in the model of dS gauge theory of gravity}

\author{Chao-Guang Huang$^a$\footnote{Email: huangcg@ihep.ac.cn},
Meng-Sen Ma$^{a,b}$\footnote{Email: mams@ihep.ac.cn}}

\medskip

\affiliation{\footnotesize $^a$ Institute of High Energy Physics, Chinese
Academy of Sciences,
Beijing 100049, China \\
$^b$ Graduate School of Chinese Academy of Sciences, Beijing, 100049, China}

\begin{abstract}

In the model of de Sitter gauge theory of gravity, the empty
homogenous and isotropic spacetimes with constant curvature scalar
and nonvanishing homogenous and isotropic torsion must have de
Sitter metrics.  The static de Sitter spacetime with static,
$O(3)$-symmetric, vector torsion is the only spherically symmetric,
vacuum solution with the metric of the form $g_{\mu\nu}={\rm
diag}(A^2(r),-B^2(r),-r^2,-r^2\sin^2\th)$.  The expressions of the
torsion for different de Sitter spacetimes are presented. They are
different from one to another. The properties of different de Sitter
spacetimes with torsion are also studied.

\end{abstract}


\maketitle

\tableofcontents
\bigskip


\section{Introduction}

The astronomical observations show that our universe is probably an
asymptotically de Sitter (dS) one \cite{SN,WMAP}. It raises
the interests on dS gauge theories of gravity.
There is a model of dS gravity{\footnote{Hereafter, the model of dS
gauge theory of gravity is called the dS gravity for short in this
paper.}}, which was first proposed in the 1970's \cite{dSG, T77}.
The dS gravity can be stimulated from dS invariant special
relativity \cite{dSSR, meetings, dSSR2} and the principle of
localization --- the full symmetry of the special relativity as well
as the laws of dynamics are both localized \cite{Guo2,vacuum,cosmos}
--- and realized in terms of the dS connection on a kind of totally
umbilical submanifolds (under the dS-Lorentz gauge) and Yang-Mills
type action \cite{dSG,Guo2,cosmos}.

It has been shown that all vacuum solutions of Einstein field
equations with a cosmological constant are the vacuum solutions of
the set of field equations without torsion \cite{Guo2, vacuum}. In
particular, Schwarzschild-dS and Kerr-dS metrics are two solutions.
On the other hand, it can also been shown that the vacuum solutions of
the set of field equations without torsion must be the vacuum
solutions of Einstein field equations with the same cosmological
constant \cite{HM}.  Therefore, one may expect that the dS gravity
may pass all solar-system-scale observations and experimental tests
for general relativity (GR)\footnote{The problem of matching the
exterior solution with an interior solution has been studied in
\cite{ZCHG}.}. It has also been shown \cite{cosmos} that the dS
gravity may explain the accelerating expansion and supply a natural
transit from decelerating expansion to accelerating expansion
without the help of the introduction of matter fields in addition to
dust.

The present paper aims at finding the dS spacetimes with torsion in
the dS gravity and studying their properties. The $k=0$ de Sitter
spacetime with constant torsion \cite{Hehl-k0dS,Minkevich} and the
static de Sitter spacetime\omits{There are some dS solutions} with
spherical torsion {which satisfy the double duality ansatz }\cite{PB,P.B} have
been presented for some gauge theories of gravity. Other de Sitter
solutions with nonvanishing
torsion are also given for other theories \cite{AP,VLS}.  But, all
of the theories  are different from the dS gravity.  We present the
de Sitter spacetimes with homogenous and isotropic torsion for
spatial curvature $k=0,\pm 1$ and static de Sitter spacetime with
static, $O(3)$-symmetric, vector torsion.  The formers are the only
vacuum solutions in the dS gravity for the empty, homogenous,
isotropic, constant-curvature-scalar universe. The latter is the
only spherically symmetric, vacuum solution in the dS gravity for a
large class of spacetimes.  \omits{the static de Sitter spacetime
with static, $O(3)$-symmetric, vector torsion is with the metric of
the form $g_{\mu\nu}={\rm diag}(A(r),B(r),r^2,r^2\sin^2\th)$.}

The paper is arranged as follows.  We first review the model of the
dS gravity in the next section. In the third section, we study the
dS solutions with homogeneous and isotropic torsion. In Sec. IV, we
solve the $O(3)$-symmetric, static , vacuum field equations. We
shall give some concluding remarks in the final section.

\section{De Sitter gauge theory of gravity}

The dS gauge theory of gravity is established based on the following
consideration.  The non-gravitational theory is de Sitter invariant special
relativity.  The theory of gravity should follow the principle of
localization, which says that the {\it full symmetry} as well as the {\it laws
of dynamics} are both localized, and the gravitational action takes
Yang-Mills-type.

A model of dS gauge theory of gravity has been constructed
\cite{dSG,T77,Guo2,vacuum,cosmos} in terms of the de Sitter connection in the
\dS-Lorentz frame, which read\footnote{The same connection with different
gravitational dynamics has also been studied (See, e.g.
\cite{MM,SW,Wil,FS,AN,Lec,Wise,Mahato,tresguerres})}
\be\label{dSc}%
({\cal B}^{AB}_{\ \ \ \,{\mu}})=\left(
\begin{array}{cc}
B^{ab}_{\ \ \,{\mu}} & R^{-1} e^a_\mu \smallskip\\
-R^{-1}e^b_\mu &0
\end{array}
\right ) \in \mathfrak{so}(1,4).
\ee%
where ${\cal B}^{AB}_{\ \ \ \, \mu}=\eta^{BC}{\cal B}^A_{\ \,
C\mu}$, in which $\eta^{AB}$ is the inverse of
$\eta_{AB}={\rm diag} (\eta_{ab}, -1)={\rm diag}(1,-1,-1,-1,-1)$ and
$e^a_{~\mu}$ is the tetrad field. Its curvature is then
\be\label{dSLF}%
 ({\cal F}^{AB}_{~~~\mu\nu})%
=\left(
\begin{array}{cc}
F^{ab}_{~~\mu\nu} + R^{-2}e^{ab}_{~~ \mu\nu} & R^{-1} T^a_{~\mu\nu}\\
-R^{-1}T^b_{~\mu\nu} &0
\end{array}
\right ) \in \mathfrak{so}(1,4),
\ee%
where $e^a_{~b\mu\nu}=e^a_\mu e_{b\nu}-e^a_\nu e_{b\mu},
e_{a\mu}=\eta_{ab}e^b_\mu$, $ F^{ab}_{~~ \mu\nu}$ and $
T^a_{~\mu\nu}$ are the curvature and torsion of the Lorentz
connection:
 \be\nno%
\vect{\Omega}^a&=&d\vect{\vartheta}^a+\vect{\omega}^a_{~b}
\wedge\vect{\vartheta}^b=\frac{1}{2}T^a_{~\mu\nu}dx^\mu\wedge
dx^\nu  \\%
&&T^a_{~\mu\nu}=\partial_\mu e^a_\nu-\partial_\nu e^a_ \mu+B^a_{~c
\mu}e^c_\nu-B^a_{~c
\nu}e^c_\mu,  \\%
\vect{\Omega}^a_{~b}&=&d\vect{\omega}^a_{~b}+\vect{\om}^a_{~c}\wedge\vect{\om}^c_{~b}=\frac{1}{2}F^a_{~b
\mu\nu}dx^\mu\wedge dx^\nu ,\nno \\ %
&&F^a_{~b
\mu\nu}=\partial_\mu B^a_{~b\nu} -\partial_ \nu
B^a_{~b\mu}+B^a_{~c\mu}B^c_{~b
\nu}-B^a_{~c\nu}B^c_{~b\mu}.  %
\ee%
where $\vect{\vartheta^a}=e^a_\mu dx^\mu$ is the coframe and
$\vect{\omega}^a_{~b}=B^a_{\ b\mu}dx^\mu$ is connection 1-form.

The action for the model of de Sitter gauge theory of gravity with
sources takes
the form of%
\be\label{S_t}%
S_{\rm T}=S_{\rm GYM}&+&S_{\mathbf M},
\ee%
where
\be\label{GYM}%
S_{\rm GYM}&=&\frac{1}{4g^2}\int_{\cal M}d^4 x e ({\cal F}^{AB}_{~~~\mu\nu}{\cal F}_{BA}^{~~~\mu\nu}) \nno \\
&=& -\int_{\cal M}d^4x e \left[{\frac{1}{4{\black g^2}}
F^{ab}_{~\mu\nu}F_{ab}^{~\mu\nu}} -{\chi(F-2\Lambda)} -
\frac{\chi}{2} T^a_{~\mu\nu}T_a^{~\mu\nu}\right]
\ee%
is the gravitational Yang-Mills action and $S_M$ is the action of
sources with minimum coupling.  In Eq.(\ref{GYM}),
$g=(R\sqrt{\chi})^{-1}\sim 10^{-61}$ is the dimensionless
gravitational coupling constant, $e=\det(e^a_\mu)$, $\La = 3/R^2$,
$\chi=1/({ 16}\pi G)$, $g^{-2}=3\chi \La^{-1}$, $G$ is the Newtonian
gravitational coupling constant, $F= -\frac{1}{2} F^{ab}_{\ \
\mu\nu}e_{ab}^{\ \ \mu\nu}$ is the scalar curvature of the Cartan
connection. ($c=1,\hbar=1$)

The field equations can be given via the variational principle with respect to
$e^a_{~\mu},B^{ab}_{~~\mu}$,
\be\label{Geq2}%
 {\cal E}_a^{\ \mu}& =&T_{a~~\ ||\nu}^{~\mu\nu } - F_{~a}^\mu
+\frac{1}{2}F e_a^\mu - \Lambda
e_a^\mu - 8\pi G( T_{{\rm M}a}^{~~\mu}+T_{{\rm G}a }^{~~\mu})=0,\qquad \\
\label{Geq2'}%
{\cal Y}_{ab}^{\ \ \mu}& =&
F_{ab~~\ ||\nu}^{~~\mu\nu} - R^{-2}(16\pi G S^{\quad \mu}_{{\rm M}ab}
+S^{\quad \mu}_{{\rm G}ab})=0.%
\ee
$||$ represents the covariant derivative defined by Christoffel symbol
$\{^\mu_{\nu\ka}\}$ and Lorentz connection $B^a_{\ b\mu}$,
$F_a^{~\mu}=-F_{ab}^{~~\mu\nu}e^b_\nu$, $F=F_a^{~\mu} e^a_\mu$,
\be T_{{\rm M}a}^{~~\mu}=-\d 1 e \d {\dl S_{\rm M}}{\dl e^a_\mu}, \qquad
S_{{\rm M}ab}^{\quad \, \mu}=\d 1 {2\sqrt{-g}}\d {\dl S_{\rm M} }{\dl
B^{ab}_{\ \ \mu}}  \ee
are the tetrad form of the stress-energy tensor and spin current for matter, respectively.
\be
T_{{\rm G}a}^{~~\mu} = \hbar g^{-2} T_{{\rm F}a}^{~~\mu}+2\chi T_{{\rm
T}a}^{~~\mu}  %
\ee
is the tetrad form of the stress-energy tensor of gravitational
field, which can be split into the curvature part
\be
\label{emF}
T_{{\rm F}a}^{~~\mu}=\omits{-\frac{1}{4e} \frac{\delta} {\delta e^a_{\mu}}\int
d^4x e {\rm Tr}(F_{\nu\ka}F^{\nu\ka}) =} e_{a}^\ka {\rm Tr}(F^{\mu \la}F_{\ka
\la})-\frac{1}{4}e_a^\mu {\rm Tr}(F^{\la \si} F_{\la \si})  \ee and torsion
part
\be%
T_{{\rm T}a}^{~~\mu}=\omits{-\frac{1}{4e} \frac{\delta} {\delta e^a_{\mu}}
\int d^4x e T^b_{\ \nu\ka}T_b^{\ \nu\ka} =} e_a^\ka
T_b^{~\mu\la}T^{b}_{~\ka\la}-\frac{1}{4}e_a^\mu T_b^{~\la\si}T^b_{~\la\si}.
\ee%
Similarly, the gravitational spin-current
\be\label{spG}%
 S_{{\rm G}ab}^{\quad \, \mu}=S_{{\rm F}ab}^{\quad \,
\mu}+2S_{{\rm
T}ab}^{\quad \,\mu} %
\ee%
can also be divided into two parts
\be%
S_{{\rm F}ab}^{\quad \, \mu}&=&\omits{\d 1 {2\sqrt{-g}}\d {\dl }{\dl
B^{ab}_{\ \ \mu}}\int d^4 x \sqrt{-g}F =}{-}e^{~~\mu \nu}_{ab\ \
{||}\nu} = Y^\mu_{~\, \la\nu}
e_{ab}^{~~\la\nu}+Y ^\nu_{~\, \la\nu } e_{ab}^{~~\mu\la},  \\
S_{{\rm T}ab}^{\quad \mu}&=& \omits{\d 1 {2\sqrt{-g}}\d {\dl }{\dl B^{ab}_{\ \
\mu}} \d 1 4 \int d^4 x \sqrt{-g}T^c_{\ \nu\la}T_c^{\ \nu\la}
=}T_{[a}^{~\mu\la}e_{b]\la}^{},
\ee%
where
\be Y^\la _{~~\mu\nu}= \d 1 2 (T^\la _{\ \,\nu\mu}+T^{\ \la} _{\mu \
\,\nu} +T^{\ \la} _{\nu \ \,\mu}) \ee
is the contortion.


\section{dS solutions with homogeneous and isotropic torsion}

First of all, there is no dS solution with SO(1,4) symmetric torsion in the
model of dS gauge theory of gravity.

For the homogeneous and isotropic universe, the metric of spacetime takes the
Friedmann-Robertson-Walker (FRW) form
 \be\label{frw}
 ds^2=dt^2-a^2(t)[\frac{dr^2}{1-k r^2}+r^2(d \th^2+ \sin^2 \th d
 \phi^2)],
 \ee
where $k=0,\pm 1$, and there are 6 Killing
vector fields $\vect{\xi}_{(I)}$ ($I=1,\cdots 6$) for each $k$.  To keep the homogeneity and
isotropy of the universe the torsion is also required to be homogeneous and isotropic.
In other words, the torsion should satisfy
\be
\mathcal{L}_{\vect{\xi}_{(I)}} \vect{T}^{a}=0, \qquad I=1,\cdots 6.
\ee
Further, we require that the torsion be invariant under space
inversion.  Then, for any $k$ the torsion always takes the form
\cite {Minkevich} \be
\begin{cases}
{\vect T}^0 = 0 & \cr %
{\vect T}^1 = {T_+}(t)\,  {\vect \vartheta}^0\wedge {\vect \vartheta}^1  & \cr %
{\vect T}^2 =  {T_+}(t)\,  {\vect
\vartheta}^0\wedge {\vect \vartheta}^2   & \cr %
{\vect T}^3 = {T_+}(t)\,  {\vect \vartheta}^0\wedge {\vect
\vartheta}^3,   &
\end{cases}
\ee %
where ${\vect \vartheta}^0=dt$, ${\vect \vartheta}^1=\frac {a(t)}{\sqrt{1-k r^2}}dr$,
${\vect \vartheta}^2=a(t) r d\th$, ${\vect \vartheta}^3=a(t) r \sin\th d\phi$.

The reduced vacuum Einstein-like equations and Yang-like equations are: %
\be \label{gee-frw-00-T+} %
&& - \d {\ddot a^2} {a^2} -  (\dot T_++ 2\d {\dot a} a  T_+ -2\d {\ddot a} a
)\dot T_+
+  T_+^4 - 4\d {\dot a} a   T_+^3 + (5 \d {\dot a^2} {a^2}   + 2\d k {a^2}- \d 3 {R^2}) T_+^2 \nno \\
&&+ 2 \d {\dot a} a(\d {\ddot a} {a} - 2 \d {\dot a^2} {a^2}  - 2  \d k {a^2}
+   \d 3 {R^2})T_+ +\d {\dot a^2}{a^2}( \d {\dot a^2}{a^2}  + 2 \d k {a^2}-
\d 2 {R^2})
+  \d {k^2}{a^4}  - \d 2 {R^2}  \d k {a^2} + \d 2 {R^4}  =0 , %
\ee
\be \label{gee-frw-11-T+} &&\d {\ddot a^2} {a^2} + (\dot T_+  + 2 \d {\dot a}
a  T_+ - 2 \d {\ddot a} a + \d {6} {R^2})\dot T_+
 - T_+^4 + 4\d {\dot a} a  T_+^3  - (5 \d {\dot a^2} {a^2} +
2 \d k {a^2}  + \d 3 {R^2}) T_+^2 \nno \\
&&- 2\d {\dot a} a (\d {\ddot a } {a}  - 2 \d {\dot a^2} {a^2 }
  - 2  \d k {a^2}- \d 6 {R^2} )T_+ - \d 4 {R^2}  \d {\ddot a} a
   - \d {\dot a^2} {a^2} (\d {\dot a^2}{a^2} +2\d  k {a^2}+   \d 2 {R^2}  )- \d {k^2}{a^4}
  -  \d 2 {R^2} \d k {a^2} + \d 6 {R^4} =0, \qquad
\ee %
\be\label{re yang}%
\ddot T_+ + 3 \d {\dot a} a \dot T_+ -( 2  T_+^2  - 6\d {\dot a} a T_+ - \d
{\ddot a} a + 5 \d {\dot a^2} {a^2} + 2 \d k {a^2}-   \d 3 {R^2}) T_+ -\d
{\dddot a} a - \d {\dot a\ddot a} {a^2} + 2 \d {\dot a^3} {a^3} + 2 \d {\dot
a} a \d k {a^2} =0 . \ee
The trace of Einstein-like equations give rise to
\begin{eqnarray}\label{dia}
\d {\ddot{a}} a =-\left (\d {\dot a} a\right )^2-\frac k {a^2} + \d 2{R^2} +
\frac 3 2 (\dot T_+ + 3\d {\dot a} a T_+- {T_+}^2).
\end{eqnarray}
A direct calculation shows the curvature scalar in this case is
\be \label{cs}
F=6\left [\d {\ddot{a}} a+\left (\d {\dot a} a\right )^2-\d
{3\dot{a}}{a}T_+-\dot{T_+}+T_+^2+\d k {a^2}\right ].
\ee
In terms of the expression, Eq.(\ref {re yang}) can be written in a very simple form
\be
 \dot{F}+2T_+(F-9/R^2)=0.
\ee
In particular, its constant-curvature scalar solutions require either $T_+=0$ or
\be \label{ccs}
F-9/R^2=0.
\ee
The former is the torsion-free case, which has been discussed in
\cite{vacuum}.  For the latter case, the combinations of Eqs.(\ref{dia}) and
(\ref{cs}) with (\ref{ccs}) give rise to
\be \label{dda}
\d {\ddot a}{a}+ \left (\d {\dot a}{a} \right)^2+\d k {a^2}=\d 1{2R^2},
\ee
\be\label{m1}
 \dot{T_+}+\d {3\dot{a}}{a}T_+-T_+^2+\d {1}{R^2}=0.
 \ee
With the help of Eqs.(\ref{dda}) and (\ref{m1}),
Eq. (\ref{gee-frw-00-T+}) reduces to 
\be\label{third} %
{\dot a}^2 +k = a^2/(4R^2). %
\ee %
Eqs.(\ref{dda}), (\ref{m1}), and (\ref{third}) constitute a (over-determinant) system
of equations for $a$ and $T_+$.

One can immediately solve Eq.(\ref{third}),
\be
a(t)=\begin{cases} H^{-1} \cosh(Ht) & \mbox{for }k=1\\
                H^{-1} e^{Ht} & \mbox{for }k=0 \\
                H^{-1} \sinh(Ht) & \mbox{for }k=-1 ,
                \end{cases}
\ee
where $H=\pm 1/2R$.  They are de Sitter solutions and satisfy
Eq.(\ref{dda}), too. The remaining task is to solve Eq.(\ref{m1})
for different de Sitter spacetimes, which can be written in
dimensionless form%
\be \label{m1'} %
y'+3\d {a'}{a} y- y^2 +4=0, %
\ee
where $y=T_+/H$, $x=Ht$, a prime represent the derivative with
respect to $x$.\\

For $k=0$ de Sitter spacetime, Eq.(\ref{m1'}) reads
\be %
y' +3 y-y^2+4=0.
\ee
It has the general solutions%
\be %
y=\d {4 + C e^{5x }} {1- Ce^{5x }} \qquad {\rm or} \qquad
T_{+}=\d {H(4 + C e^{5Ht })}{1- Ce^{5Ht }},
\ee %
where $C$ is a constant of integration.  In particular, when
$C=\infty$, $y=-1$, $T_+=\mp \frac 1 {2R}$. When $C=0$, $y=4$,
$T_+=\pm \frac 2 R$. Both of them are constant solutions.  They are
also the asymptotical states at $x\to \pm \infty$ for a generic $C$.
When $x\to  -\frac 1 5 \ln C$, $y$ and thus $T_+$ tends to $\infty$.
Fig. 1 plots the dimensionless torsion $y$ versus the dimensionless
time $x$ for $C=1$.
\begin{figure}[th]
\centerline{\includegraphics[width=9cm]{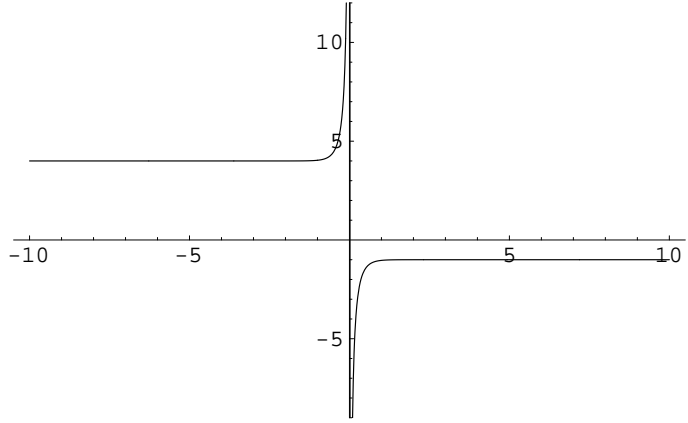}} \caption{The
evolution of dimensionless torsion $y$ for $k=0$ de Sitter spacetime
and $C=1$.}
\end{figure}

The stress-energy tensors and the spin currents of gravitational
field are
\be
 T_{{\rm F}\ b}^{\ a}&=&12H^2(T^2_{+}-2HT_{+}-2H^2){\rm diag}(3, -1, -1,
-1) \medskip \nno \\
&=& 12H^4\d {(6+18Ce^{5Ht}+C^2e^{10Ht})}{(1-Ce^{5Ht})^2}{\rm
diag}(3, -1, -1,
 -1),  \medskip \\
 T_{{\rm T}\ b}^{\ a} &=& \d {T^2_{+}}{2}{\rm diag}(3, -1, -1,
-1) 
=
\d {H^2}{2}\d {(4+Ce^{5Ht})^2}{(1-Ce^{5Ht})^2}{\rm diag}(3, -1,
-1,-1), \medskip \\
S_{{\rm F}ab}^{\ \ \ c} &=&  4 S_{{\rm T}ab}^{\ \ \ c} =-2T_{+}
(\dl^c_a \dl^0_b-\dl_a^0\dl_b^c) 
=
-2H\d {4 + C e^{5Ht}}{1- Ce^{5Ht}} (\dl^c_a
\dl^0_b-\dl_a^0\dl_b^c),
\ee%
respectively.  In particular, they reduce to
\be%
&&T_{{\rm F}\ b}^{\ a}= \d {\La^2} {12} {\rm diag}(3, -1, -1, -1) \\
&& T_{{\rm T}\ b}^{\ a}=\d {\La} {24} {\rm diag}(3, -1, -1, -1) \\
&&S_{{\rm F}ab}^{\ \ \ c} =  4S_{{\rm T}ab}^{\ \ \ c} =\pm
\sqrt{{\La}/ {3}} (\dl^c_a \dl^0_b-\dl_a^0\dl_b^c)
\ee%
for the case $C=\infty$, $T_+=-H=\mp\frac 1 2 \sqrt{\La
/3}$, \omits{the stress-energy tensors and the spin currents are,
respectively,}and
\be%
&&T_{{\rm F}\ b}^{\ a}=\frac{\La^2}{2} {\rm diag}(3, -1, -1, -1) \\
&& T_{{\rm T}\ b}^{\ a}= \frac{2\La}{3} {\rm diag}(3, -1, -1,-1) \\
&&S_{{\rm F}ab}^{\ \ \ c} =  4 S_{{\rm T}ab}^{\ \ \ c} =\mp
4\sqrt{{\La}/ {3}} (\dl^c_a \dl^0_b-\dl_a^0\dl_b^c)
\ee%
for the case $C=0$, $T_+=4H=\pm 2\sqrt{\La /3}$, \omits{the stress-energy
tensors and the spin currents are, respectively,}which are all finite everywhere.  For other $C$, the stress-energy
tensors and the spin currents will be divergent at $t=- \frac 1 {5H}
\ln C$.

For the $k=+1$ de Sitter spacetime, Eq.(\ref{m1'}) reads
\be%
y'(x)+ 3 \tanh(x) y-y^2=-4  .%
\ee%
FIG 2 presents the finite, non-oscillatory, numerical solutions
$y(x)$ with $T_+ \sim o(H)$.
\begin{figure}[th]
\centerline{\includegraphics[width=9cm]{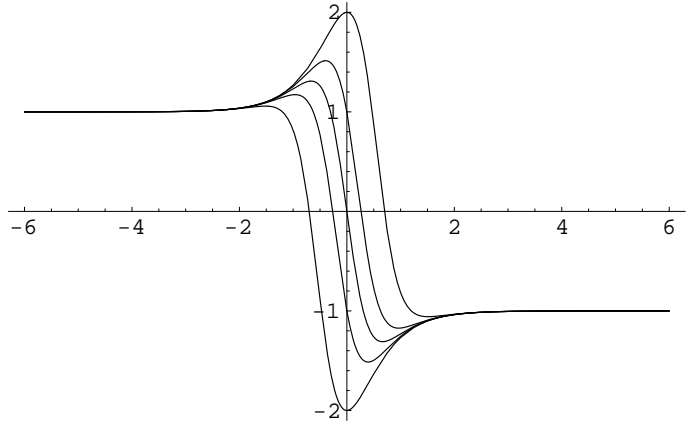}}
\caption{Finite, non-oscillatory, numerical solution $y(x)$ with
$T_+ \sim o(H)$.}
\end{figure}

The stress-energy tensors of gravitational fields
\be%
&&T_{{\rm F}\ b}^{\ a}= {12H^2}[T_+^2-2H \tanh(Ht)T_+-2H^2]
 {\rm diag}(3, -1, -1, -1),  \\
&&T_{{\rm T}\ b}^{\ a}= \d {T_+^2} {2} {\rm diag}(3, -1, -1, -1),
\ee%
and the spin currents of gravitational fields
\be%
S_{{\rm F}ab}^{\ \ \ c} =  4S_{{\rm T}ab}^{\ \ \ c} =-2T_+
(\dl^c_a \dl^0_b-\dl_a^0\dl_b^c),
\ee%
are all finite everywhere.

For the  $k=-1$ de Sitter spacetime, Eq.(\ref{m1'}) reduces to
\be\label{Riccati2}%
y'(x)+ 3\coth(x) y-y^2=-4 , %
\ee%
 which has the asymptotical solution $y= -1$ as
$x\to \infty$ and the asymptotical solution %
\be%
y \to 2 {\rm csch}^2  x \{\coth x+[\log(\tanh \frac x 2)-C] \sinh x \}^{-1}%
\ee%
when $x\to 0$.  Therefore, $T_+(t)$ should be initially huge and
decay to $T_+=-H$ as $t\to \infty$.  FIG 3 sketches out the numerical
solutions for Eq.(\ref{Riccati2}).
\begin{figure}[th]
\centerline{\includegraphics[width=9cm]{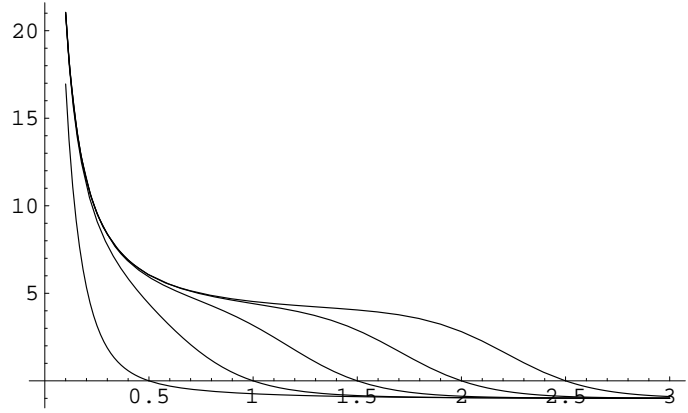}}
\caption{The numerical solution for Eq.(\ref{Riccati2}).}
\end{figure}

In this case, the stress-energy tensors of
gravitational fields
\be%
&&T_{{\rm F}\ b}^{\ a}= {12H^2}[T_+^2-2H \coth(Ht)T_+-2H^2]
{\rm diag}(3, -1, -1, -1),   \\
&& T_{{\rm T}\ b}^{\ a}= \d {T_+^2} {2} {\rm diag}(3, -1, -1, -1),
\ee%
and the spin currents of gravitational fields
\be%
&&S_{{\rm F}ab}^{\ \ \ c} = 4S_{{\rm T}ab}^{\ \ \ c} =-2T_+ (\dl^c_a
\dl^0_b-\dl_a^0\dl_b^c), \
\ee%
are all initially divergent.

\section{dS solutions with static, spherically symmetric torsion}
To find the static dS solutions we suppose that the metric takes the form
\be \label{static}%
ds^2=A^2(r)dt^2 -B^2(r)dr^2 -r^2 d\Om^2.%
\ee
On the same reason as the cosmological case above, the torsion
should be static and O(3) invariant, namely, \be {\cal
L}_{\vect{\xi}_{(I)}}\vect{T}^a=0,\qquad I=1,\cdots 4 \ee where
$\vect{\xi}_{(I)}$ ($I=1\cdots 4$) are the timelike Killing vector
fields and three rotation Killing vector fields and $\vect{T}^a$ is
invariant under the space inversion. {Generally the static spherically symmetric torsion can
be taken as the forms in the papers \cite{Nester}}. Further, the torsion can be
irreducibly decomposed as trace-vector, axial-vector and tensor
pieces under the Lorentz group.  For static and O(3)-symmetric
torsion, the axial-vector piece automatically vanishes.  For
simplicity, we consider the trace-vector piece
\be \label{statictorsion}%
{\vect T}^0 &=& T_0(r) {\vect \vartheta}^0\wedge {\vect \vartheta}^1 \nno \\
{\vect T}^1 &=& T_1(r) {\vect \vartheta}^0\wedge {\vect \vartheta}^1 \nno \\
{\vect T}^2 &=& T_1(r) {\vect \vartheta}^0\wedge {\vect \vartheta}^2
-T_0(r){\vect \vartheta}^1\wedge {\vect \vartheta}^2 \nno \\
{\vect T}^3 &=& T_1(r){\vect \vartheta}^0\wedge {\vect
\vartheta}^3-T_0(r) {\vect \vartheta}^1\wedge {\vect
\vartheta}^3,%
\ee%
 where ${\vect \vartheta}^0=A(r)dt$, ${\vect \vartheta}^1=B(r)dr$,
${\vect \vartheta}^2=rd\th$, and ${\vect \vartheta}^3=r\sin \th d\phi$.

By substituting the coframe and torsion forms into
Eqs.(\ref{Geq2}) and (\ref{Geq2'}) we can get 9 independent
equations, 5 for the Einstein-like equation and 4 for the Yang-like
equation, which are listed in Appendix for completion. Below we can
simplify the field equations to get the de Sitter solutions.

{Firstly ${\cal E}_1^{\ \mu}e^0_\mu - {\cal E}_0^{\ \mu}e^1_\mu=0$ ( i.e. (\ref{E10}) $-$
(\ref{E01}))} gives rise to%
\be\label{AT1}%
A(r)T_1(r)=\bar C.
\ee%
where $\bar C$ is an arbitrary constant with  the dimension of the
inverse of length.

{Then with the help of Eq.(\ref{AT1}), ${\cal Y}_{01}^{\ \ \mu}e^1_\mu +
{\cal Y}_{20}^{\ \ \mu} e^2_\mu=0$ (i.e. (\ref{Y011}) $+$ (\ref{Y202}))} leads to%
\be \label{T1''}%
T_1'' = T_1' (\d {2T_1'} {T_1} + \d {B'} B + \d 1 r)+\d 1 r T_1(\d {B'} B + \d
{1-B^2} r  ).%
\ee%
Additionally one can obtain {from
${\cal Y}_{01}^{\ \ \mu}e^1_\mu=0$ (i.e.  Eq.(\ref{Y011}))}%
\be \label{T0'}%
T_0' = \d 1 {rB} (\d {T_1'} {T_1}+\frac {B'} {B}) +
\frac{T_{0}}{T_{1}} T_1'+
 B (T_1^2 - T_0^2- \d \La 2 ) -\d 2
r T_0.%
\ee%
The trace of Einstein-like equations, {namely ${\cal E}_a^{\ \mu}
e^a_{\mu}=0$} then gives rise to
\be \label{T1'}%
\d {T_1'} {T_1} = -\d {B'} B + \d \La 6 r B^2 .
\ee%
Then, the system of equations
Eqs.(\ref{E00})---(\ref{Y122}) reduce to
\omits{,independent equations become}%
the differential equation%
\be\label{RDEq}%
2r \d {B'} B - \d \La 4 r^2 B^2  + B^2 - 1=0 %
\ee%
and algebraic equation%
\be\label{Beq}%
\La r^2 B^2 - 12 B^2 + 12=0
\ee%
plus Eqs.(\ref{AT1}) --- (\ref{T1'}). From
Eq.(\ref{Beq}), one immediately finds%
\be \label{B}%
B^2=\d 1 {1- H^2 r^2},%
\ee%
where $H^2=\La/12$.  It also solves Eq.(\ref{RDEq}) obviously.
The integration of Eq.(\ref{T1'}) shows%
\omits{\be%
\d{{T_1^2}'} {2T_1^2}&=&\d 1 {2r}(- \d \La 4 r^2 B^2  + B^2 - 1) + \d \La 6 r
B^2\nno \\
&=&\d {B^2} {2r}(- \d \La 4 r^2   + 1- \d 1 {B^2} ) + \d \La 6 r B^2 \nno\\
&=&\d {B^2} {2r}(- \d \La 4 r^2   + \d \La {12}r^2 ) + \d \La 6 r B^2\nno\\
&=& \d \La {12} r B^2%
\ee}%
\be\label{T1}%
T_1^2 = \d {C_1^2H^2} {1- H^2r^2}.%
\ee%
where $C_1$ is a dimensionless, integration constant.
Eq.(\ref{T1}) is consistent with Eq.(\ref{T1''}) as it should be.  Further,%
\be%
A^2=\d {\bar C^2} {H^2C_1^2} (1-H^2 r^2).%
\ee%
Without loss of generality, one can choose $\bar C^2= H^2 C_1^2 = H^2 C^2$ by
the re-scale of the time
$t$.  Then,%
\be\label{A}%
A^2= (1-H^2 r^2).%
\ee%
Eq.(\ref{T0'}) becomes
\be\label{T0'new}%
T_0'+\d {2-3H^2r^2}{r(1-H^2r^2)} T_0+\d
{1}{(1-H^2r^2)^{1/2}}T_0^2 = -  \d {4H^2}
{(1-H^2r^2)^{1/2}} +\d {C^2H^2}{(1-H^2 r^2)^{3/2}}.%
\ee%
It can be written as%
\be\label{statict0}%
\d {dy}{d\zeta}+\frac{2-\tan^2 \zeta}{\tan \zeta} y + y^2 ={C^2} \sec^2\zeta-4,%
\ee%
where $y =T_0/H$, $r=H^{-1}\sin \zeta$. The  general solution of the
equation is a  function of hypergeometric functions  with an
integration of constant $C_0$. The reality of both $y$ and  $C_{0}$
 requires  $C_{0}$ to be
zero. Thus, the solution takes the form
\be \label{gsol}%
y(\zeta)  =\left [4
        - \d { C^2-16} 7 \sec^2\zeta \d {F(\frac {6-C} 2,\frac {6+C}2,\frac{9}{2},
        \sec^2\zeta)} {F(\frac {4-C} 2,\frac {4+C}
        2,\frac{7}{2},\sec^2\zeta)}
         \right ]  \tan\zeta .
\ee%
As a special case, $C=4$, the solution reduces to
\be y(\zeta) & =&4\tan\zeta \quad
\mbox{or}\quad T_{0}=\frac{4H^2r}{\sqrt{1-H^2r^2}} , %
\ee
which has the asymptotic behavior
\be%
 T_0 \to \d {CH}
{\sqrt{1-H^2r^2}}=T_1{\yellow .}
  \ee%
as $\zeta \to \pi/2$ or $r \to H^{-1}$. In fact, the general
solution  (\ref{gsol}) also  shares the same asymptotic property.
The behavior of $T_0(r)$  is shown in FIG 4, which is not sensitive
to $T_0(0)$.
\begin{figure}[thb]
\centerline{\includegraphics[width=9cm]{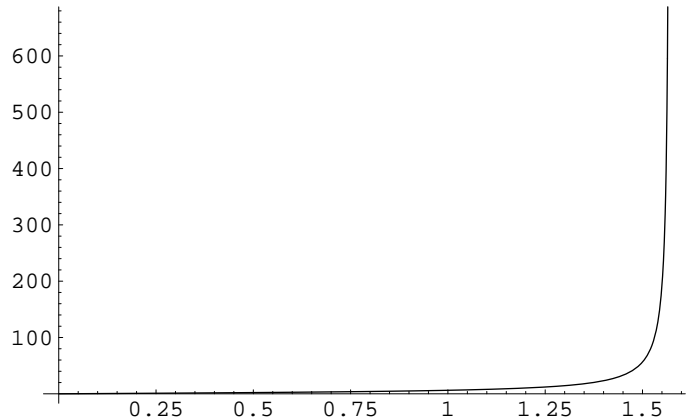}}
\caption{ $y$ thus $T_0$ is divergent at $\zeta=\pi/2$ or
$r=H^{-1}$.}
\end{figure}

In brief, the static dS space
\be%
ds^2=(1-H^2 r^2)dt^2 -\d {dr^2} {1-H^2 r^2} -r^2 d\Om^2 \nno%
\ee%
with the static torsion $T_0$ and $T_1$, given by Eqs.(\ref{gsol})
and (\ref{T1}), respectively, is the  only
solution of the vacuum field equations for the Ansatz (\ref{static}) and (\ref{statictorsion}).%

The nonzero tetrad components of the stress-energy tensor of gravitational
fields are
\be%
&&\left \{ \begin{array}{l}
T_{{\rm F}\ 0}^{\ 0}= \La [T_0^2+T_1^2+ 2H^2 -\d {2H^2r}{\sqrt{1-H^2r^2}}T_0],  \\
T_{{\rm F}\ 0}^{\ 1}= -T_{{\rm F}\ 1}^{\ 0}=2T_1\La [T_0-\d {H^2r}{\sqrt{1-H^2r^2}}], \\
T_{{\rm F}\ 1}^{\ 1}= \La [T_1^2-3T_0^2 -6H^2 -\d 2 r \d
{2-3H^2r^2}{\sqrt{1-H^2r^2}}T_0], \\
T_{{\rm F}\ 2}^{\ 2}= T_{{\rm F}\ 3}^{\ 3}=\La [T_0^2-T_1^2+ 2H^2 +\d 2 r
T_0\sqrt{1-H^2r^2}],
\end{array} \right . \\
&&\left \{ \begin{array}{l}
T_{{\rm T}\ 0}^{\ 0}= \frac 1 2 (T_0^2+3T_1^2), \\
T_{{\rm T}\ 0}^{\ 1}= -T_{{\rm T}\ 1}^{\ 0}=2T_0 T_1, \\
T_{{\rm T}\ 1}^{\ 1}= -\frac 1 2 (3T_0^2+T_1^2), \\
 T_{{\rm T}\ 2}^{\ 2}= T_{{\rm T}\ 3}^{\ 3}=\frac 1 2 (T_0^2-T_1^2). \end{array} \right .
\ee%

The spin currents of gravitational fields are
\be%
S_{{\rm F}ab}^{\ \ \ c} &=&4S_{{\rm T}ab}^{\ \ \ c} = -2T_1 (\dl^c_a
\dl^0_b-\dl_a^0\dl_b^c)+2T_0 (\dl^c_a \dl^1_b-\dl_a^1\dl_b^c),\
\ee%
The straightforward calculation shows that
\be%
F_{a\mu}F^{a\mu}&=&4\left (3T_0^4+T_1^4-4T_0^2T_1^2+\d 4 r
\d{2-3H^2r^2}{\sqrt{1-H^2r^2}}T_0^3- \d 4 r\d
{1-2H^2r^2}{\sqrt{1-H^2r^2}}T_0T_1^2\right . \nno \\
&&\left . +\d 2 {r^2}\d {3-2H^2r^2}{1-H^2r^2}T_0^2-
2H^2\d{2-H^2r^2}{1-H^2r^2}T_1^2+\d
{8H^2}{r}\d{2-3H^2r^2}{\sqrt{1-H^2r^2}}T_0 +93H^4 \right ) \omits{\nno \\
&\to& 4\left (3T_0^4+T_1^4-4T_0^2T_1^2-
\d{4HT_0(T_0^2-T_1^2)}{\sqrt{1-H^2r^2}}+\d
{2H^2(T_0^2-T_1^2)}{1-H^2r^2} \right . \nno \\
&&\left . -\d{8H^3T_0}{\sqrt{1-H^2r^2}}+93H^4\right ) \to \infty },
\ee%
\omits{\be%
4H^2(2\d {(1-2C)(T_0^2-T_1^2)-4H^2C}{1-H^2r^2} +93H^2) \ee}
\be%
\hspace{-1.5cm}F_{a\mu}F_{b\nu}e^a_\nu e^b_\mu &=&4\left \{\d
{H^4r^4}{(1-H^2r^2)^2}(\sqrt{1-H^2r^2}-3) T_0^4
+3T_0^4+T_1^4-4T_0^2T_1^2\right .\nno \\
&&+\d 4 r\sqrt{1-H^2r^2}[2T_0^2-(1-2H^2r^2)T_1^2]T_0\nno \\
&& +\d 4 {r\sqrt{1-H^2r^2}}[{2H^2(2-3H^2r^2)}- {H^2r^2} T_0^2]T_0\nno \\
&&+\d 2 {1-H^2r^2}[\d {3-2H^2r^2} {r^2}T_0^2-H^2(2-H^2r^2)T_1^2] \left .-93 \d
{H^4}{1-H^2r^2} \right \}.
\ee%
Obviously, they are divergent when $r\to 1/H$.

\section{Concluding remarks}

 The vacuum equations of the dS gravity have been solved and several important dS
solutions have been obtained.  The $k=0$ dS spacetime with
constant torsion and $k=+1$ dS spacetime have no singularity in
whole spacetimes. In contrast, the $k=0$ dS spacetime with varying
torsion, $k=-1$ dS spacetime, and the static dS spacetime have
singularity.

{The most important feature of our
solutions is that the different de Sitter metrics describe different
geometries because of the existence of the nonzero torsions and their dependence on different
coordinates in different manners. Although the metric admits 10 Killing vector fields
for each case, the torsion does not possess so high symmetry.  For
different metrics, the torsion has different symmetry. Even for the
3 homogeneous and isotropic cases, the symmetries are different.
When $k=0,\pm 1$, the torsion is ISO(3)-, SO(4)-, SO(3,1)-symmetric,
respectively. Therefore, they are not equivalent to each other.

Another important feature of our solutions is that for the static de Sitter spacetime
with torsion the horizon is no longer a coordinate singularity, Since the torsion,
gravitational stress-energy tensor, and invariant curvature scalars become divergent at singularities,
these singularities are intrinsic ones in the dS gravity.} In GR
the horizon singularity is a kind of coordinate singularity which can
be removed by coordinate transformation. Here
these singularities at horizon are not coordinate singularities and
can not be removed. so the Riemann-Cartan spacetime can not be
extended to pass through the horizon. Some properties about horizon
in GR, like Hawking radiation and horizon entropy, should be
reconsidered in this kind of theory of gravity.

\begin{acknowledgments}\vskip -4mm
This work is supported by NSFC under Grant Nos. 10775140 and Knowledge
Innovation Funds of CAS (KJCX3-SYW-S03).
\end{acknowledgments}

\begin{appendix}
\section{Field equations for Ansatz (4.1)
and (4.2)} The explicit expressions of the field equations for
Ansatz (\ref{static}) and (\ref{statictorsion}) are as follows.  The
independent Einstein-like equations $(T_{a\ \ \, ||\nu}^{\
\mu\nu}e^b_\mu +\cdots)$ are ($ab:$) $00$, $ 10$, $ 01$, $11$ and
$22$ component equations.  They are
\be \label{E00}%
&& - 3 \d {A''} A \left [\d 1 2 \d {A''}  A - \d {A'} A \d {B'}
B+ B\left ( \d {A'} A  T_0+T_0' \right ) \right ] + 3\d {(A')^2}
{A^2} \left ( B^2 T_1^2 -\d
1 2 \d {(B')^2} {B^2}-  \d 1 {r^2}\right) \nno \\
&& + 3\d {(A')^2} {A^2} \left (B' -
 \d 3 2 B^2  T_0 - \d 2 r B  \right)T_0 + 3\d {A'} A  \left (B' T_0'-\d 2 {r^2} BT_0\right ) \nno\\
& & - 3\d {A'} A B^2T_0 \left (T_0' + 2 B  T_0^2 - 2 B T_1^2+\d 4 r
T_0\right ) -\d 1 {r}\left (6B' T_0'- \d 3 r \d {(B')^2} {B^2}
 + 2\La B {B'}  \right)\nno\\
 &&+ \d 6 {r}B' \left (B T_1^2 -\d 1 r T_0\right )  + \frac 3 2 B^{2}T_0'\left (T_0'  - 4 B T_1^2+
 \d 4 r T_0+ 2 \La B \right ) -
 3 B^{2}T_1' ( T_1' - 2 B T_0 T_1)\nno \\
 && +\frac 3 2B^4\left(3 T_1^4 -  T_0^4 - 2 T_0^2 T_1^2  + \La  T_0^2- 3\La  T_1^2- \d 2 {r^{2}} T_0^2
 + \d 2 {r^2} T_1^2 +\d 1
 {r^{4}}  -  \d {2\La} {3r^{2}}+
  \d 2 3\La^2 \right) \nno\\
 &&- \d 6 rB^{3}  T_0 \left (2  T_1^2 + \d 1 {r^2}- \La\right ) +
\d 1 {r^2}B^{2}\left( 9 T_0^2 - 3
 T_1^2 - \d 3 {r^2} + \La \right ) + \frac 3 {2r^3}\left( 4B T_0 +\d 1 r \right )=0,\qquad %
\ee%
\be \label{E10}%
&& \frac{A'}{A} \left ( B(T_1T_0'   - T_0T_1')-\d 1 r T_1'  + {B}^2T_1({T_0}^2-
{T_1}^2 ) +  \d 2 r {B}T_1T_0-
 \d 1 r \d {B'} B T_1\right)- \frac 1 2 \La{B}^2 T_1'\nno \\
     && +{B}T_0
  \left( {B} (T_1 T_0' - T_0T_1')-\d 1 r T_1'   +
  {B}^2T_1({T_0}^2 - {T_1}^2 +\frac 1 2 \La)+
  \frac{2}{r}B T_0T_1  - \d 1 r\frac{B'}{B}T_1\right)  =0, \qquad
\ee%
\be\label{E01}%
\left (B T_0 + \d {A'} A \right) \left (B (T_1 T_0'- T_0 T_1')-
   \d 1 r T_1' + B^2 T_1 (T_0^2 -  T_1^2+\d \La 2 ) +
     \d 2 r B T_0 T_1   -
   \d 1 r \d {B'} B T_1
  \right) =0, %
\ee%
\be\label{E11}%
&&\hspace{-1cm}- \frac 1 2 \d {A''^2}{A^2}+ \d {A''} A\left(\d {A'} A(\d
{B'} B -  B T_0) - BT_0' \right ) -  \d {A'^2} {A^2} \left (\d 1 2
\d {B'^2}{B^2}-T_0 B'-B^2(\d 1 2T_0^2 - T_1^2)-\d 2 r
B T_0 - \d 1 {r^2} \right )  \nno\\
&& +   B \d {A'} A \left( (\d {B'} B T_0' - BT_0 T_0'+\d 2 {r^2}
T_0) + 2B^2 T_0 (T_0^2 - T_1^2 +\d \La  2)+ B(
\d {4} r T_0^2 + \d {2\La} {3r})\right ) - \d 1 {r^2} \d {B'^2} {B^2}     \nno \\
&& +\d 2 r B'(T_0'- B T_1^2+ \d 1 r T_0 )  + B^2 \left( T_1'^2- \d 3
2 T_0'^2
+2BT_1 (T_1 T_0'- T_0T_1') - \d {2} r T_0 T_0'\right )\nno \\
&&+B^4\left( \d 3 2 T_0^4 - \d 1 2  T_1^4 - T_0^2 T_1^2- \d 1
{r^2}(T_0^2-T_1^2) + \d \La  2 (3T_0^2 - T_1^2) +
\d 1 {2r^4} -  \d {\La} {3r^2} + \d 1 3 \La^2  \right)\nno \\
&&+ \d 2 r B^3  T_0 \left ( 2 T_0^2+ \La - \d 1 {r^2} \right )+\d 1
{r^2} B^2 \left( {3}  T_0^2 -  T_1^2+ \d \La {3}-\d 1 {r^2}\right )
+\d 1 {r^3} (2B T_0 +  \d 1 {2r})
  =0,%
\ee%
\be\label{E22}%
&& \frac 1 2 \d {A''^2}{A^2}+ \d {A''} A\left(\d {A'} A(B T_0
- \d {B'} B ) + BT_0' +\d \La 3 B^2\right ) +  \d {A'^2} {A^2} \left
(\d 1 2 \d
{B'^2}{B^2}-T_0 B'+\d 1 2B^2 T_0^2\right )  \nno\\
&& - B \d {A'} A \left( (\d {B'} B T_0'-\d \La 3 B'- BT_0 T_0'
) -\La B^2 T_0 -\d {\La} {3r}B\right ) -\d \La 3 \d 1 r B B'   +
B^2T_0' \left( \d 1 2 T_0'
+\La B \right )\nno \\
&&+B^4\left( -\d 3 2 (T_0^4 + T_1^4) + T_0^2 T_1^2+ \d 1
{r^2}(T_0^2-T_1^2) + \d \La  2 (T_0^2 - 3T_1^2) +
\d 1 {2r^4}  + \d 1 3 \La^2  \right)\nno \\
&&+ \d 2 r B^3  T_0 \left (T_1^2-  T_0^2+\d \La 2+ \d 1 {r^2} \right
)-\d 1 {r^2} B^2 \left( {3}  T_0^2 -  T_1^2-\d 1 {r^2}\right ) -\d 1
{r^3} (2B T_0 + \d 1 {2r})
  =0.%
\ee%
The independent Yang-like equations
$(F_{ab\ \ \,||\nu}^{\ \ \mu\nu}e^c_\mu+\cdots )$ are ($abc:$)
$010$, $011$, $202$ and $122$ component equations.%
\be\label{Y010}%
&&\d {A'''} A - \d {A''} A\left(\d {A'} A + 3 \d {B'} B- B
T_0- \d 2 r\right )- \d {B''} B \d {A'} A + \d {A'^2}{A^2}\left (\d
{B'} B - B T_0 \right)+ \d {A'}A \d {B'} B\left (3 \d
{B'} {B} -  B T_0 - \d 2 r\right ) \nno \\
&& + \d {A'}A \left (BT_0'  - 2  B^2 T_0^2 + 2  B^2 T_1^2 -  \d 2 r
B T_0 - \d 2
{r^2} \right ) + BT_0''- B' T_0' + \d 2 r BT_0'\nno \\
&& + 2  B^3 T_0 \left (T_1^2-   T_0^2-  \d \La 2 \right )
 - \d 2 {r}B T_0\left (2B  T_0  + \d 1 r\right )=0,%
\ee%
\be\label{Y011}%
 2B^2(T_1' T_0 - T_0' T_1) + \d 2 r (BT_1' + B' T_1)
  + 2 B^3 T_1 (T_1^2-  T_0^2-\d 1 2 \La)  - \d 4 r B^2 T_0 T_1=0,%
\ee%
 \be\label{Y202}%
&&B\left (\d {A'^2} {A^2} T_1 - \d {A'} A T_1'
  + \d {B'} B  T_1'
    -  T_1''
 +\d 1 {r^2} T_1 (1-B^2)\right ) + B^2 T_0 (3\d {A'} A T_1+ T_1' )
 \nno \\
&&+ 2B^2  T_0'  T_1-\d 1 r (B'T_1+ BT_1') - 2B^3T_1  (T_1^2-T_0^2 -\d 1 2 \La)
+  \d 4 r B^2 T_0T_1 =0,%
\ee%
\be\label{Y122}%
&&\d {A'^2} {A^2} (B T_0 + \d 1 r) - \d {A'}A(B T_0' - 2 B^2
  T_0^2- B^2T_1 ^2+ \d 1 r B T_0- \d 1 r \d {B'} B)
 +  \d 1 r(\d {B''} B - 3 \d {B'^2}{B^2}) \nno \\
&& -  (BT_0'' - B' T_0' - \d 1 r B' T_0 +\d 2 r BT_0') + 3B^2T_1
T_1'
  + B^3 T_0( 2  T_0^2 - 2  T_1^2
  - \d 1 {r^2} +\La  ) \nno \\
&&  +  \d 1 r B^2 (4T_0^2 -  \d 1 {r^2}) +   \d 3 {r^2} BT_0
 + \d 1 {r^3}=0.%
\ee%

\end{appendix}


\begin{thebibliography}{07}
\bibitem{SN}
A. G. Riess \textit{et al}., Astron. J. {\vect 116} (1998), 1009,
astro-ph/9805201; \omits{
\bibitem{SN2}
Supernovae Cosmology Project Collaboration,} S. Perlmutter
\textit{et al}., Astrophys. J. {\vect 517} (1999), 565,
astro-ph/9812133. \omits{
\bibitem{SN3}}
A. G. Riess \textit{et al}., Astrophys. J. {\vect 536} (2000), 62,
astro-ph/0001384. \omits{
\bibitem{SN4}
Supernova Search Team Collaboration,} A. G. Riess \textit{et al}.,
Astrophys. J. {\vect 560} (2001), 49, astro-ph/0104455.

\bibitem{WMAP}
C. L. Bennett \textit{et al}., Astrophys. J. Suppl. {\vect 148}
(2003), 1, astro-ph/0302207; \omits{ \ddot{}\bibitem{WMAP2} WMAP
Collaboration,} D. N. Spergel \textit{et al}., Astrophys. J. Suppl.
{\vect 148} (2003), 175, astro-ph/0302209.

\bibitem{dSG}Y.-S. Wu, G.-D. Li and H.-Y. Guo, Kexue Tongbao (Chi. Sci. Bull.)
{\vect 19} (1974), 509; Y. An, S, Chen, Z.-L. Zou and H.-Y. Guo,
{\it ibid}, 379; H.-Y. Guo, {\it ibid} {\vect 21} (1976) 31;  Z. L.
Zou,  et al,  Sci. Sinica {\vect XXII} (1979), 628; M.-L. Yan, B.-H.
Zhao and H.-Y. Guo, \omits{Renormalization of gravitation field with
torsion,} Kexue Tongbao (Chi. Sci. Bull.) {\vect 24} (1979), 587;
{\it Acta
Physica Sinica} {\vect 33} (1984), 1377; 1386 (all in Chinese). 


\bibitem{T77} P. K. Townsend, \PRD15 (1977), 2795; A. A. Tseytlin, \PRD26
(1982), 3327.



\bibitem{dSSR} K.-H. Look (Q.-K. Lu) 1970, Why the Minkowski metric must be
used? unpublished; K.-H. Look, C.-L. Tsou (Z.-L. Zou) and H.-Y. Kuo
(H.-Y. Guo),  Acta Phys. Sin. {\vect 23} (1974), 225;  Nature
(Shanghai, Suppl.); H.-Y. Guo, Kexue Tongbao (Chinese Sci. Bull.)
{\vect 22} (1977), 487 (all in Chinese).

\bibitem{meetings}\omits{A
theory of inertial motion in space-time of constant curvature,}H.-Y.
Guo, in {\it Proceedings of the 2nd Marcel Grossmann Meeting on
General Relativity}, ed. by R. Ruffini, (North-Holland, 1982), 801;
\omits{The meaning of relativity in spacetimes of constant
curvature,} {\it Nucl. Phys. B} (Proc. Suppl.) {\vect 6} (1989),
381; C.-G. Huang and H.-Y. Guo, in {\it Gravitation and Astrophysics
--- On the Occasion of the 90th Year of General Relativity,
Proceddings of the VII Asia-Pacific International Conference}, ed.
by J. M. Nester, C.-M Chen, and J.-P. Hsu (World Scientific, 2007,
Singapore), 260.

\bibitem{dSSR2}
H.-Y. Guo, C.-G. Huang, Z. Xu and B. Zhou,\omits{On Beltrami model
of de Sitter spacetime, arXiv:} {\it Mod. Phys. Lett.} {\vect A19}
(2004), 1701, 
hep-th/0403013;
\PLA331 (2004), 1, hep-th/0403171; H.-Y. Guo, C.-G. Huang, Z. Xu and
B. Zhou, {\it Chin. Phys. Lett.} {\vect 22} (2005), 2477,
hep-th/0508094;
 H.-Y. Guo, C.-G. Huang,  Y. Tian, Z. Xu and B. Zhou,
 {\it Acta Phys. Sin.} {\vect 54} (2005), 2494 (in Chinese); 
 M.L. Yan, N.C. Xiao, W. Huang, S. Li, {\it Comm. Theor. Phys.} {\vect48} (2007) 27;53;
 H.-Y. Guo, C.-G. Huang and B. Zhou, {\it Europhys. Lett.} {\vect 72} (2005),
 1045, hep-th/0404010; Z. Chang, S. X. Chen, C.-G. Huang,
Chin. Phys. Lett. 22 (2005), 791 54;  Z. Chang, S.-X. Chen, C.-B. Guan, and
C.-G. Huang, 
Phys. Rev. D 71 (2005), 103007;  S.-X. Chen, N.-C. Xiao, and
M.-L.Yan, 
hep-th/0703110.

\bibitem{Guo2}
H.-Y. Guo, C.-G. Huang, Y. Tian, H.-T. Wu, B. Zhou, Class. Quant.
Grav. {\vect 24} (2007) 4009; H.-Y. Guo, {\it Phys. Lett.} {\vect
B653} (2007) 88; H.-Y. Guo, Scin. Chin. arXiv:0707.3855

\bibitem{vacuum} H.-Y. Guo, C.-G. Huang, Y. Tian, B. Zhou, Front. Phys.
China, {\vect 2} (2007), 358.

\bibitem{cosmos} C.-G. Huang, H.-Q. Zhang, H.-Y. Guo,
JCAP 10 (2008) 010;
Chin. Phys. C (High Energ. Phys. Nucl. Phys.) 
{\vect 32} (2008), 687.

\bibitem{HM}C.-G. Huang and M.-S. Ma, in preparation.

\bibitem{ZCHG} Z.-L. Zou, {\it et al}, {Acta Astron. Sinica} {\vect 17} (1976) 147; S. Chen,
{\it et al}, {Scientia Sinica} (1976), 35 (both in Chinese).



{\bibitem{Hehl-k0dS}P. Baekler and F. W. Hehl, in Gauge Theory
and Gravitation
--- Proceedings of the International Symposium on Gauge Theory and Gravitation
(g \& G), ed K. Kikkawa, N. Nakanishi, and H.Nariai,
(Springer-Verlag, Berlin 1983) 1.

\bibitem{Minkevich} A. V. Minkevich, Phys. Lett. A {\vect 80} (1980), 232; {\vect 95} (1983),
422; F. M\"uller-Hoissen, Phys. Lett. A {\vect 92} (1982), 433.}

\bibitem{PB} P. Baekler, F. W. Hehl, and E.W.Mielke, \omits{Vacuum solutions with double duality properties of
the Poincar\'e gauge field theory II}in General Relativity ---
Proceedings of the 2nd Marcel Grossmann Meeting on the Recent
Progress of General Relativity, ed R. Ruffini, (North-Holland,
Amsterdam, 1982), 413; P. Baekler, F. W. Hehl, and H.-J. Lenzen,
\omits{Vacuum solutions with double duality properties of the
Poincar\'e gauge field theory II}in General Relativity ---
Proceedings of the 3rd Marcel Grossmann Meeting on the Recent
Progress of General Relativity, ed N. Hu, (Science Press, Beijing
and North-Holland, Amsterdam, 1983), 107.


\bibitem{P.B} P. Baekler, Phys. lett. B {\vect 99} (1981), 329;
Phys. lett. A {\vect 96} (1983), 279.

\bibitem{AP} R. Aldrovandi and J. G. Pereira, An Introduction to Teleparllel Gravity
(2001$\sim$2007). %

\bibitem{VLS} de Vega, Larsen, and S\'anchez, Phys. Rev. D {\vect 58} (1998) 026001;

\bibitem{MM} S. W. MacDowell and F. Mansouri, \PRL\ {\vect 38}
(1977), 739; Erratum-ibid. {\vect 38} (1977), 1376.

\bibitem{SW} K. S. Stelle and P. C.
West, \PRD21 (1980), 1466.

\bibitem{Wil}F. Wilczek, \PRL\ {\vect 80} (1998), 4851.

\bibitem{FS} L. Freidel and A. Starodubtsev, Quantum gravity in terms of topological observables, arXiv: hep-th/0501191.

\bibitem{AN}J. Armenta and J. A. Nieto,
J. Math. Phys. {\vect 46} (2005), 112503.

\bibitem{Lec}M. Leclerc, Annals of Physics {\vect 321} (2006),
708.

\bibitem{Wise} D.K. Wise, MacDowell-Mansouri Gravity and Cartan
Geometry, arXiv: gr-qc/0611154.


\bibitem{Mahato}P. Mahato, 
 Mod. Phys. Lett. A {\vect 17} (2002) 1991-1998, arXiv:gr-qc/0604042;
Phys. Rev. D{\vect 70} (2004) 124024, arXiv:gr-qc/0603100;
Int. J. Theor. Phys. {\vect 44} (2005) 79-93, arXiv:gr-qc/0603109;
Int. J. Mod. Phys. A {\vect 22} (2007), 835, arXiv:gr-qc/0603134.

\bibitem{tresguerres}R.~Tresguerres, Int. J. Geom. Meth. Mod. Phys. {\vect 5}
(2008), 171.

{\bibitem{Nester} R.-S. Tung, C.-H Chang, D.-C Chern, J.M Nester, Progress of Theoretical Physics,
{\vect 88} (1992), 291; J.-K.Ho, D.-C. Chern, J.M Nester, Chinese Journal of Physics, {\vect 35} (1997),640.}




\omits{\bibitem{PGT} T. W. B. Kibble, {\it J. Math. Phys.} {\vect 2}
(1961), 212; F. W. Held, P. von der Heyde, G. D. Kerlick, and
J. M.  Nester,  
 {\it Rev Mod. Phys.} {\vect 48} (1976), 393 and references therein.

\omits{\bibitem{Wise} D.K. Wise, MacDowell-Mansouri Gravity and Cartan
Geometry, arXiv: gr-qc/0611154.}
\bibitem{dSconnection}  S. W. MacDowell and F. Mansouri, \PRL\ {\vect 38} (1977), 739;
Erratum-ibid. {\vect 38} (1977), 1376;
K. S. Stelle and P. C. West, \PRD21 (1980), 1466;
F. Wilczek, \PRL\ {\vect 80} (1998), 4951;
L. Freidel and A. Starodubtsev, Quantum gravity in terms of topological
observables, arXiv: hep-th/0501191;
M. Leclerc,  Annals of Physics, {\vect 321} (2006), 708; E. Witten,
Three-Dimensional Gravity Reconsidered, arXiv:0706.3359.

\omits{
\bibitem{gtg79}
 H.Y. Kuo,
 in {\it Proc.   2nd M. Grossmann Meet. on GR} (1979), ed. by R. Ruffini,
 North-Holland Publ. (1982) 475.}

\bibitem{Yang} C. N. Yang, Phys. Rev. Lett. {\vect 33} (1974), 445.

\bibitem{wzc}Y.-S. Wu, Z.-L. Zou and S. Chen, {\it Kexue Tongbao (Chin. Sci.
Bull.)} {\vect 18} (1973), 119 (in Chinese).

\bibitem{GD} H. Stephani, D. Kramer, M. MacCallum, C. Hoenselaers, E.
Herlt, Exact Solutions of Einstein's Field Equations, (Cambridge University
Press, Cambridge, 1980).

\bibitem{RauchNieh} R. Rauch and H. T. Nieh, Phys. Rev. D{\vect 24} (1981), 2029.
}

\end{thebibliography}
\end{document}